\pdfoutput=1
\documentclass[12pt]{iopart}
\usepackage{graphicx}
\usepackage{iopams}
\usepackage{xspace}
\usepackage{verbatim}

\begin{document}

\title{Superfluid Bloch dynamics in an incommensurate optical lattice}

\author{J B Reeves$^1$, B Gadway$^1$ \footnote{\address{Present address: JILA, National Institute of Standards and Technology and University of Colorado, Department of Physics, University of Colorado, Boulder, CO 80309-0440, USA}}, T Bergeman$^1$, I Danshita$^{2,3}$ and D Schneble$^1$}
\address{$^1$ Department of Physics and Astronomy, Stony Brook University, Stony Brook, New York 11794-3800, USA}
\address{$^2$ Yukawa Institute for Theoretical Physics, Kyoto University, Kyoto University, Kyoto 606-8502, Japan}
\address{$^3$ Computational Condensed Matter Physics, RIKEN, Wako, Saitama 351-0198, Japan}
\ead{jeremy.reeves@stonybrook.edu}

\begin{abstract}
We investigate the interplay of disorder and interactions in the accelerated transport of a Bose-Einstein condensate through an incommensurate optical lattice. We show that interactions can effectively cancel the damping of Bloch oscillations due to the disordered potential and we provide a simple model to qualitatively capture this screening effect. We find that the characteristic interaction energy, above which interactions and disorder cooperate to enhance, rather than reduce, the damping of Bloch oscillations, coincides with the average disorder depth. This is consistent with results of a mean-field simulation.

\end{abstract}

\maketitle

The combined effects of disorder and interactions in condensed-matter systems can influence their transport properties in profound ways. Beyond merely reducing conductivity, disorder can lead to Anderson localization \cite{Anderson1958} in the absence of interactions, and repulsive interactions can, in turn, give rise to localized Mott phases \cite{Hubbard1963} without the influence of disorder. While disorder and interaction act cooperatively in the limit of strong interactions by promoting disordered insulating phases \cite{Giamarchi1988,Fisher1989}, weak interactions in bosonic systems can counteract the localizing effects of disorder by making accessible higher-energy states, thus screening the disorder potential \cite{Scalettar1991,Deissler2010}.  The often subtle interplay between disorder and interactions is relevant for condensed-matter systems such as superfluid helium in porous media and disordered films \cite{Reppy1992, Crowell1997} and granular superconductors \cite{Azbel1980,Shklovskii2007}, but it can also be studied experimentally with ultracold atoms in optical lattices. While optical lattices are naturally defect free \cite{Lewenstein2012}, it is possible to implement disorder using laser speckle \cite{Billy2008,Kondov2011}, localized impurity atoms \cite{Gadway2011}, and incommensurate standing waves \cite{Lye2005}. Recent experiments with disordered optical potentials have addressed Anderson localization \cite{Billy2008,Kondov2011,Roati2008}, delocalization due to repulsive interactions \cite{Deissler2010,Lye2007,Kondov2013}, as well as the formation of glassy, localized phases in the regime of strong repulsive interactions \cite{Gadway2011,Krinner2013a,White2009,Fallani2007,Tanzi2013}.

Disorder and interactions also fundamentally affect the dynamics of particles in lattice potentials.  Static external forcing results in coherent Bloch oscillations (BOs) \cite{Bloch1929, Zener1934} that prevent macroscopic transport unless there exist mechanisms for relaxation \cite{Esaki1970}. While BOs are overdamped in conventional solids, they can be observed in superlattice semiconductors \cite{Waschke1993} and also with ultracold atoms in optical lattices \cite{Dahan1996,Anderson1998,Morsch2006}. When using atomic quantum gases, minimizing the collisional interactions allows for the direct observation of extremely long-lived BOs \cite{Roati2004, Fattori2008, Gustavsson2008}. Typical mean-field interactions in Bose-Einstein condensates already result in significant dephasing \cite{Fattori2008, Gustavsson2008} due to nonlinearities, dynamical instability, and quantum chaotic dynamics \cite{Witthaut2005, Kolovsky2009, Buchleitner2003} (it has recently been observed that in certain parameter regimes, such dephasing can be reversible \cite{Kolovsky2003,Meinert2013}). A qualitatively similar dephasing of BOs in optical lattices was also observed to occur due to the additional presence of disorder, which scrambles the regular phase evolution between lattice sites \cite{Drenkelforth2008}. 

For the case that both interactions and disorder are present, it has been predicted that the combined effects can modify the Bloch oscillation dynamics, either reducing or enhancing the damping due to disorder, depending on their relative magnitude \cite{Schulte2008, Walter2010}. In the present manuscript, we study the interplay between disorder and interactions in the dynamics of Bloch oscillations in an incommensurate lattice. Our results demonstrate that for a given disorder-induced damping rate, increased interactions can reduce or enhance the overall damping rate. Furthermore, we identify a characteristic interaction strength between the regimes of reduction and enhancement, corresponding to a typical disorder energy scale. We provide a simple model to capture the basic features of the system.


In our experiment we load a BEC of a few $10^4$ $^{87}$Rb atoms in the  $|F,m_F\rangle \equiv |2,-2 \rangle$ hyperfine state into a crossed optical dipole trap (XODT) \cite{Pertot2009}. A magnetic field gradient levitates the BEC in the trap, before a one-dimensional optical lattice
\begin{equation}
V(z)=s_1 E_R \cos^2 (k_1 z) + s_2 E_R \cos^2 (k_2 z) 
\label{Lattice Potential}
\end{equation}
is adiabatically ramped up along the vertical direction $\hat{z}$. Here $k_j=2 \pi/\lambda_j$, and $s_1$ ($s_2$) are the lattice depths in units of the main lattice recoil energy $E_R = (\hbar k_1)^2/(2m)$, $\hbar=h/(2\pi)$ is Planck's reduced constant and $\lambda_1=1064$ nm and $\lambda_2=785.2$ nm are the wavelengths of the lattice.  The ratio $\beta=k_2/k_1 \approx 1.3551$ is used as a measure of the lattice commensurability; for nearly irrational $\beta$, the lattices will be periodic on length scales exceeding those relevant to the experiment. For all measurements, the main lattice has a depth of $s_1=3$ and the second lattice depth is varied from $s_2=0$ to $s_2 \approx 1$. We define the disorder strength $\Delta$ as the average energy shift of a lattice site due to the secondary lattice, $\Delta \approx s_2/2$, which also corresponds to the energy scale $E_\textrm{\tiny{gap}}\approx s_2/2 E_R$ of perturbations to the band structure.

\begin{figure}[t!]
    \centering
    \includegraphics[height=2.5in]{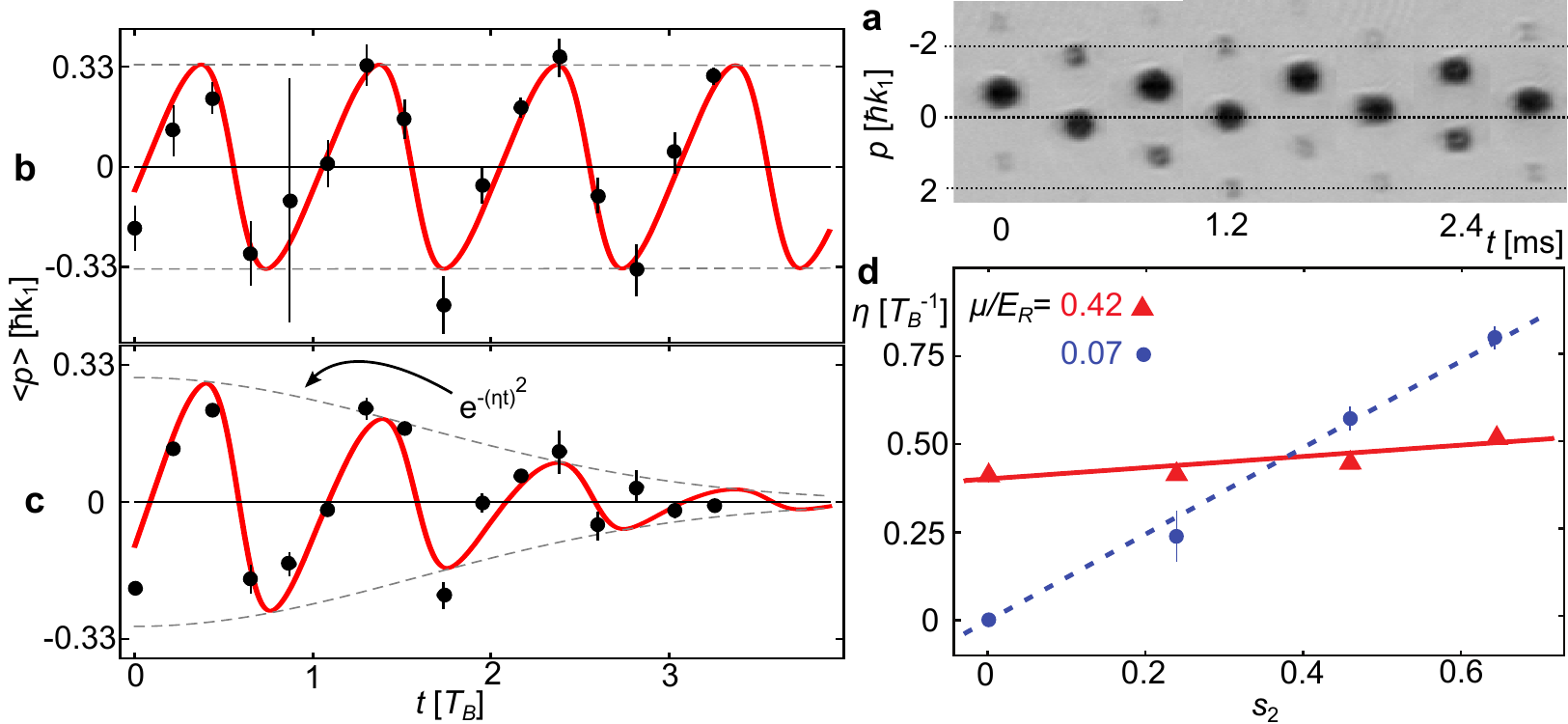}
\caption{Bloch oscillations of weakly interacting atoms in an optical lattice.
(a) Absorption images for different evolution times in a single-wavelength lattice. (b,c) Observed average momentum for a weakly interacting cloud ($\tilde{\mu}=0.07$) for $s_2=0$ and 0.46 respectively. The red lines are fits to the data while the dashed grey lines trace the Gaussian envelope with decay rate $\eta$. The extracted damping rates in b) and c) are $0.02$ ${T_B}^{-1}$ and $0.55$ ${T_B}^{-1}$ respectively.  (d) Damping rate as a function of secondary lattice depth for $\tilde{\mu}=0.07$ (blue, dashed, circles) and for $\tilde{\mu}=0.42$ (red, solid, triangles). The lines are linear fits to guide the eye. In (b-d), each data point represents three repetitions of the experiment, error bars represent standard deviations.
}
    \label{FIG:BOs}
\end{figure}

The chemical potential of the condensate is adjusted in a range $\tilde{\mu}=\mu/E_R=0.1...0.5$ prior to loading the lattice by varying the atom number as well as the  trapping frequency of the XODT from $\sim 20$ Hz to $\sim 80$ Hz. To induce BOs the levitating gradient is rapidly switched off in $< 200$ $\mu$s, and the atoms are allowed to evolve in the lattice plus XODT potentials for a variable hold time. During this time the atoms undergo BOs with period $T_B=h/Fd \approx 0.9$ ms where $F=mg$ is the gravitational force, and $d=\lambda_1/2$ is the main lattice spacing; for our system $Fd=0.56$ $E_R$. The atoms are then suddenly released and the atomic momentum distribution is imaged on a CCD camera after 18~ms time of flight, as shown in Figure~\ref{FIG:BOs} (a).


After repeating the experiment for different hold times, the oscillations are recorded as seen in Figure~\ref{FIG:BOs} (b,c). A damping rate $\eta$ is extracted from the data by fitting the mean of the linear momentum with 
\begin{equation}
\langle p_z(t) \rangle \propto e^{-(\eta t)^2} \frac{dE(q)}{dq} \frac{dq}{dt}.
\label{FitFnt}
\end{equation}
Here $E(q)$ is the energy dispersion of the first band of the lattice as a function of quasimomuntum $q$ which evolves as $\hbar q=F t$. The Gaussian envelope is chosen because at short times the $p_z(t)\approx (1-t^2)$ behaviour best matches dephasing due to two beating frequencies \cite{Walter2010}. This also matches previously observed and simulated BO damping behaviour due to both interactions and incommensurate lattices \cite{Meinert2013,Walter2010}, while a pure exponential envelope has been suggested in the disorder-free case \cite{Buchleitner2003}.

In the weakly interacting ($\tilde{\mu}=0.07$) and disorder-free case (Figure~\ref{FIG:BOs} (b)) we observe several 10 $T_B$ of only weakly damped oscillations, with damping due to the nonzero interaction energy and the harmonic confinement of the XODT.  When adding the second lattice with $s_2=0.46$ the damping time is markedly reduced to just a few $T_B$, as seen in Figure~\ref{FIG:BOs} (c). As illustrated in Figure~\ref{FIG:BOs} (d), increasing the secondary lattice depth increases the damping rate for a weakly interacting cloud; however, as shown for two interactions $\tilde{\mu}=0.07$ and 0.42, the effect of the disorder is weaker when stronger interactions are present. Furthermore, the effect of interactions depends on the depth of the disorder potential: for small $s_2<0.35$, the damping rate is higher for low $\tilde{\mu}$ while the opposite is true for $s_2>0.35$.


\begin{figure}[t!]
    \centering
    \includegraphics[height=2.5in]{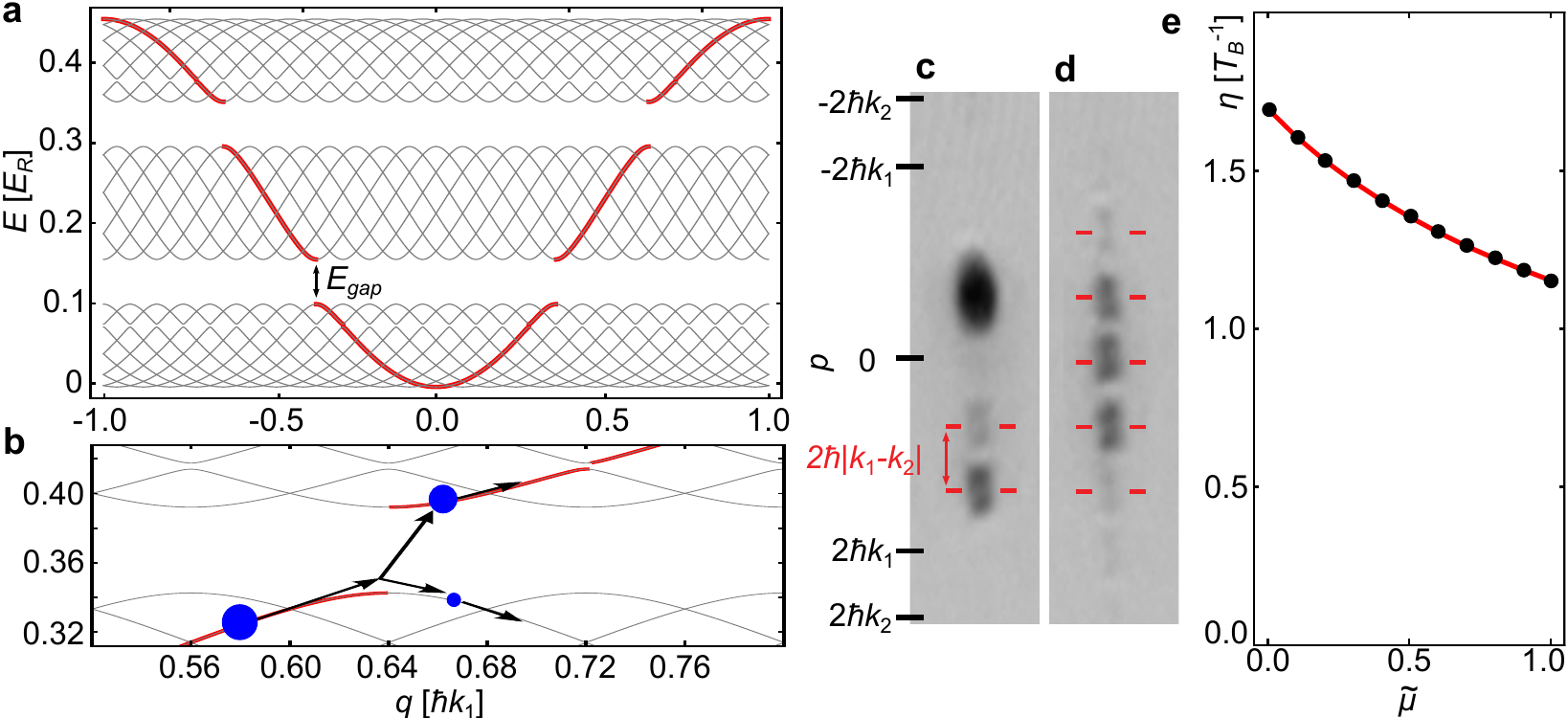}
\caption{Damping due to the band structure of the bichromatic lattice.
 a) Band structure of the lowest energy band of the two-colour lattice calculated in the tight-binding limit for $s_1=3$ and $s_2=0.1$.  Our irrational $\beta=\lambda_1/\lambda_2$ is approximated by $\beta=1.36$, which is still nonperiodic on experimental length scales. The bold red curve outlines the perturbed lowest band of the $s_2=0$ lattice while the grey lines represent minibands of the bichromatic potential.  The primary effect of the second lattice is to open minigaps in the lowest band. b) Schematic illustration of a Landau-Zener tunnelling event at a mini gap. (c,d) Momentum distributions of atoms undergoing oscillations in a band structure described in a) after 0.65 $T_B$ and 2.4 $T_B$ respectively. The red lines represent a relative momentum of $2\hbar|k_1-k_2|$.  The damping rate due to the minigaps for $s_2$=0.4 is plotted in e) for a range of $\tilde{\mu}$ as calculated using a nonlinear Landau-Zener formalism (see text).
}
    \label{FIG:LZ}
\end{figure}


To elucidate this further, we investigate the damping in the noninteracting and interacting regimes separately. In the noninteracting case, it can be described by considering the single particle band structure of the lattice potential. Starting with all the atoms at $q=0$ in the lowest band, for $s_2=0$ the atoms experience a continuous, periodic dispersion relation $E(q)$ giving rise to long-lived Bloch oscillations described by (\ref{FitFnt}).  The chief effect of the second lattice, seen in Figure~\ref{FIG:LZ} (a), is to open minigaps of energy $E_\textrm{\tiny gap}$ in the band structure at $k_2$ and $|k_1-k_2|$ \cite{Diener2001}. The minigaps serve to disrupt the evolution of the momentum distribution by splitting the condensate among the minibands of the lattice through Landau-Zener tunnelling \cite{Zener1934}, as illustrated in Figure~\ref{FIG:LZ} (b). Because the minigaps appear at irrational fractions of the original Brillouin zone width 2$\hbar k_1$, the rephasing time for the split condensate is much longer than the oscillation lifetime in our experiment.

The fraction of atoms which tunnel to those that pass the minigap adiabatically is approximately 
\begin{equation}
P=\exp{\Big [}\frac{-2 \pi E^2_\textrm{\tiny gap}}{ F (\rmd E/\rmd q|_\textrm{\tiny gap})}{\Big ]}.
\label{LZeq}
\end{equation}
where $\rmd E/\rmd q|_\textrm{\tiny gap}$ is the unperturbed band's slope at the gap. In absorption images of the momentum distribution, the splitting is clearly seen after passing through just one half of the Brillouin zone, Figure~\ref{FIG:LZ} (c). Unlike in Figure~\ref{FIG:BOs} (a), where a single colour lattice gives peaks separated by $2\hbar k_1$, we observe an extra peak at a momentum separation of $2\hbar|k_1-k_2|$.  For longer evolution times as in Figure~\ref{FIG:LZ} (d), the condensate breaks up into a series of peaks with spacing $2\hbar|k_1-k_2|$, spanning a range of momenta between $\pm 2 \hbar k_1$.  This leads to a breakdown of collective oscillation and accelerated damping. 

Mean-field interactions in the BEC modify the damping by changing the miniband shape near the minigap, causing the minibands to develop a swallow-tail structure,  corresponding to a discontinuous, multivalued energy spectrum \cite{Wu2000}. The discontinuity leads to the breakdown of adiabaticity in the strong-interaction limit $\mu \gtrsim E_\textrm{\tiny gap}$, resulting in an expected increase in the fraction $P$ of atoms that tunnel through the gap and remain on the perturbed original band. The effect of the nonlinearities can also be considered as an effective rescaling of the gap size in (\ref{LZeq}) according to the interaction energy. The effective $\Delta_\textrm{\tiny{gap}}$ is as much as 50\% smaller for $\mu \approx \Delta_\textrm{\tiny{gap}}$.

To estimate the damping of BOs due to these mechanisms, we numerically propagate a wavepacket through the bichromatic dispersion relation. For simplicity, the calculation does not keep track of the accrued phases. Additionally, at times when a wavepacket reaches a minigap, it is instantaneously split according to a nonlinear Landau-Zener calculation in which interactions have the effect of increasing the tunnelling rate without explicitly including the swallow-tail structure in $E(q)$.  The results, shown for $s_2=0.4$ in  Figure~\ref{FIG:LZ} (e), predict a decrease in the damping rate as the interactions are ramped up. We note that this simple calculation does not separately address the damping induced by interactions themselves, which perturb the evenly spaced Wannier-Stark energy levels of the tilted(forced)-lattice, giving rise to irreversible damping for our lattice parameters in which the energy offset between lattice sites is comparable to the tunnelling matrix element \cite{Buchleitner2003,Meinert2013}.

The predicted decrease of the Landau-Zener induced damping due to interactions as in Figure~\ref{FIG:LZ} (e) agrees with our experimental data to within a factor of 3. The discrepancy could be due to a number of factors.  For example, our calculation does not take into account the three dimensional nature of the experimental setup nor does it include possible coherence effects for the Landau-Zener tunnelling, which are difficult to incorporate in the quasiperiodic potential. The assumption that the tunnelling through a minigap happens all at once could also alter the calculated dynamics as the condensate has a finite momentum width and passes through many minigaps in rapid succession.

\begin{figure}[t!]
    \centering
    \includegraphics[height=2.5in]{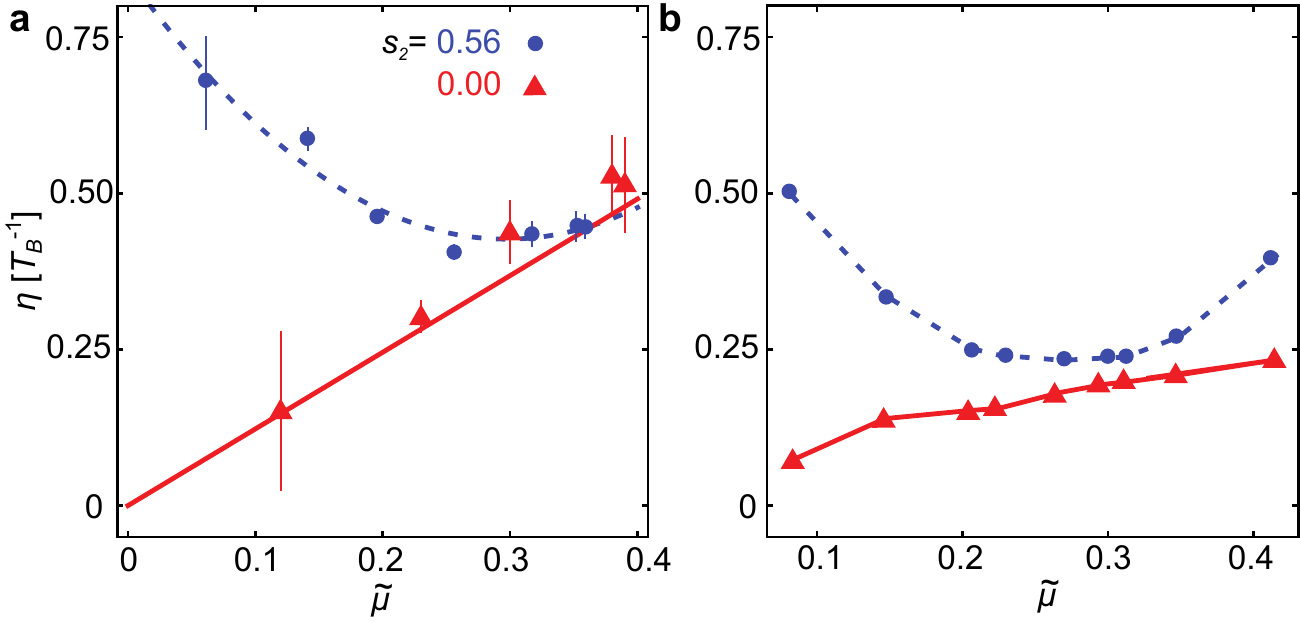}
\caption{Damping rates of BOs as a function of chemical potential.  (a) Measured damping rate as a function of $\tilde{\mu}$ for different $s_2$. For the $s_2=0$ case (red,solid, triangles), the damping rate increases roughly linearly with $\tilde{\mu}$. For a larger disorder $s_2=0.56$ (blue, dashed, circles) a clear decrease in the damping rate is observed as the $\tilde{\mu}$ is increased. Lines are fits to guide the eye.  From a parabolic fit to the $s_2=0.56$ data a minimum at $\tilde{\mu}_c= 0.29$ is obtained. In b) we plot the same data obtained from a 3DGPE simulation.  A fit finds a minimum damping rate at $\tilde{\mu}_c=0.26$. 
}
    \label{FIG:Dip}
\end{figure}

Experimentally we have selected our parameters to best probe the relationship between interaction-induced damping and Landau-Zener damping due to the potential's quasi-disorder.   Figure~\ref{FIG:Dip} (a) plots the measured damping rate with varied chemical potential for different disorder depths, as described in Figure~\ref{FIG:BOs} (b,c). In the $s_2=0$ case we observe, as expected from earlier related work \cite{Gustavsson2008}, a steady increase in the interaction-induced damping. For $s_2>0$, the initial damping in the weakly interacting case is significantly enhanced. However, for increasing interactions, within an experimentally accessible range, we subsequently  observe a \emph{decrease} in the damping rate, only to be followed later by a small increase.  A heuristic parabolic fit yields a chemical potential $\tilde{\mu}_{co}=0.29$ for minimum damping.

The observed behaviour is consistent with a simple qualitative picture in which the interactions initially screen out the disorder, thus allowing for longer lived oscillations.  Above $\tilde{\mu}_{co}$ the damping driven by interactions outweighs the screening and the damping rate begins to increase. We note that $\tilde{\mu}_{co} \approx \Delta = 0.28$, indicating that the crossover from disorder dominated to mean-field dominated behaviour occurs when the interacton strength becomes comparable to the disorder. This is consistent with findings for the equilibrium case \cite{Deissler2010}, where a weakly interacting condensate with $\tilde{\mu} <\Delta$ tends to become fragmented,  corresponding to a breakdown of global coherence across the cloud, and where the fragmentation becomes less severe leading to increased phase coherence as the potential is ``filled'' by the interaction energy. Above $\tilde{\mu}_{co}$,  the amount of fragmentation is limited with only weak dependence $\tilde{\mu}$ and mean-field effects take over as the disorder is of minimal significance to the cloud's coherence.

To confirm our experimental finding that $\mu_{co} \simeq \Delta$, we perform a three dimensional Gross-Pitaevskii equation (3DGPE) simulation with experimental parameters as inputs (see Figure~\ref{FIG:Dip} (b)). Data from the simulation is analysed in a manner identical to the experimental data and a very similar behaviour of the damping rate is observed, with a minimum at  $\tilde{\mu}_{co}=0.26$  in good agreement with our experimental findings.  We note that the factor of $\sim 1.3$ absolute discrepancy between the experimental and simulated rates could result from the fact that our full 3DGPE does not account for depletion of the condensate.

In summary, we have demonstrated that interactions and disorder can compete to determine the transport properties of a lattice trapped superfluid. The competition results in the decrease of the disorder-induced damping rate of BOs.  Our findings for Bloch oscillations are also consistent with other recent transport measurements in disordered potentials, including dipole oscillations in a quasiperiodic potential \cite{Lye2007} and in a 2D speckle potential \cite{Krinner2013}. The observed effect saturates at a minimum damping rate when the interaction energy becomes comparable to the characteristic energy scale of the disorder.

The interaction-induced delocalization effects observed here are also relevant for the description of systems ranging from superfluid helium in porous materials to superconductivity in granular or disordered materials \cite{Reppy1992, Crowell1997,Azbel1980,Shklovskii2007}.  Here the competition between disorder and ineractions ultimately contrasts with cooperative behaviour in the regime of stronger interactions, which can give rise to strongly localized, disordered phases such as the Bose glass \cite{Giamarchi1988,Fisher1989}. Our findings in the crossover between the two regimes give an example of how subtle the parameter dependence can be in a given system.

\ack

The authors thank Matthias Vogt for contributions to experimental work. We acknowledge support from NSF (PHY-0855643 and PHY-1205894),  as well as NSF PHY-0968905 (T.B.), and the GAANN program of the U.S. DoEd (J.R., B.G.).  I.D. acknowledges support from KAKENHI (grants No. 25800228 and 25220711).

\end{document}